\documentclass[twocolumn,aps,prc,superscriptaddress,showpacs,floatfix,longbibliography]{revtex4-1}
\usepackage{url}
\usepackage{cancel}
\usepackage[colorlinks,linkcolor=blue,citecolor=blue,filecolor=black,urlcolor=blue]{hyperref}
\usepackage{epsfig,graphics}
\usepackage{graphicx}
\usepackage{dcolumn}
\usepackage{bm}
\usepackage[usenames]{color}
\usepackage{amssymb}
\usepackage{amsmath}
\usepackage{multirow}
\usepackage{float}
\usepackage{harpoon}
\usepackage{MnSymbol}
\usepackage{appendix}
\usepackage{color}
\usepackage{hyperref}
\usepackage{cleveref}

\newcommand{\nch} {N_{\mathrm{ch}}}

\newcommand{\pT} {p_{\mathrm{T}}}
\newcommand{\lr}[1]{\left\langle #1\right\rangle}

\newcommand{\pbpb}{$^{208}$Pb+$^{208}$Pb}
\newcommand{\auau}{$^{197}$Au+$^{197}$Au}

\newcommand{\ruru}{$^{96}$Ru+$^{96}$Ru}
\newcommand{\zrzr}{$^{96}$Zr+$^{96}$Zr}

\usepackage{CJK}
\hfuzz=\maxdimen
\begin{document}
\title{Precision tests of the nonlinear mode coupling of anisotropic flow\\ via high-energy collisions of isobars}
\newcommand{\sbu}{Department of Chemistry, Stony Brook University, Stony Brook, NY 11794, USA}
\newcommand{\bnl}{Physics Department, Brookhaven National Laboratory, Upton, NY 11976, USA}
\newcommand{\itp}{Institut f\"ur Theoretische Physik, Universit\"at Heidelberg, Philosophenweg 16, 69120 Heidelberg, Germany}
\author{Jiangyong Jia}\email[Correspond to\ ]{jiangyong.jia@stonybrook.edu}\affiliation{\sbu}\affiliation{\bnl}
\author{Giuliano Giacalone}\email[Correspond to\ ]{giulianogiacalone@gmail.com}\affiliation{\itp}
\author{Chunjian Zhang}\affiliation{\sbu}
\begin{abstract}
Valuable information on the dynamics of expanding fluids can be inferred from the response of such systems to perturbations in their initial geometry. We apply this technique in high-energy $^{96}$Ru+$^{96}$Ru and $^{96}$Zr+$^{96}$Zr collisions to scrutinize the expansion dynamics of the quark-gluon plasma, where the initial geometry perturbations are sourced by the differences in deformations and radial profiles between $^{96}$Ru and $^{96}$Zr, and the collective response is captured by the change in anisotropic flow $V_n$ between the two collision systems. Using a transport model, we analyze how the nonlinear coupling between lower-order flow harmonics $V_2$ and $V_3$ to the higher-order flow harmonics $V_4$ and $V_5$, expected to scale as $V_{4\mathrm{NL}}=\chi_4 V_2^2$ and $V_{5\mathrm{NL}}=\chi_5 V_2V_3$, gets modified as one moves from $^{96}$Ru+$^{96}$Ru to $^{96}$Zr+$^{96}$Zr systems. We find that these scaling relations are valid to high precision: variations of order 20\% in $V_{4\mathrm{NL}}$ and $V_{5\mathrm{NL}}$ due to differences in quadrupole deformation, octupole deformation, and nuclear skin modify $\chi_{4}$ and $\chi_5$ by about 1--2\%. Percent-level deviations are however larger than the expected experimental uncertainties and could be measured. Therefore, collisions of isobars with different nuclear structures are a unique tool to isolate subtle nonlinear effects in the expansion of the quark-gluon plasma that would be otherwise impossible to access in a single collision system.
\end{abstract}

\maketitle
\paragraph*{Introduction.} The space-time evolution of the quark-gluon plasma (QGP) produced in high-energy nuclear collisions is driven by pressure-gradient forces that convert spatial deformations in the initial geometry into momentum-space deformations that are captured by azimuthal correlations among final-state hadrons \cite{Ollitrault:1992bk,Alver:2010gr,Teaney:2010vd}. These correlations emerge in the Fourier spectrum of the azimuthal particle distribution, $p(\phi)\propto \sum_{n=-\infty}^{\infty}V_n e^{in\phi}$ \cite{Heinz:2013th}. The dissipative effects in the QGP expansion quickly dampen the $V_n$ spectrum as one goes higher in $n$~\cite{Teaney:2012ke}. The dominant coefficients, reflecting genuine deformations of the QGP geometry, are elliptic flow, $V_2$, and triangular flow, $V_3$. Damped higher harmonics are strongly affected by their couplings to $V_2$ and $V_3$.  Quadrangular flow, $V_4$, for instance, receives a large contribution from its coupling to $V_2$, which scales as $V_2^2$.

Over the past decade, many theoretical \cite{Gardim:2011xv,Teaney:2012ke,Teaney:2013dta,Gardim:2014tya,Yan:2015jma,Qian:2016fpi,Giacalone:2016afq,Qian:2016pau,Giacalone:2016mdr,Qian:2017ier,Giacalone:2018wpp,Magdy:2021sba,Zhao:2022uhl} and experimental \cite{ATLAS:2014ndd,ATLAS:2015qwl,ALICE:2017fcd,ALICE:2020sup,STAR:2020gcl} investigations have clarified the mechanism by which $V_2$ and $V_3$ source harmonics of higher order. The resulting picture is that $V_{n,n>3}$ contains a contribution reflecting a genuine $n^{\mathrm{th}}$-order deformation, denoted as $V_{n\mathrm{L}}$,  plus a contribution from couplings to $V_2$ and $V_3$, denoted as $V_{n\mathrm{NL}}$. For $V_4$ and $V_5$, the decompositions are~\cite{Yan:2015jma},
\begin{align}
\label{eq:1}
V_{4}=V_{4 \mathrm{L}}+\chi_{4}\left(V_{2}\right)^{2}\;, \hspace{20pt}   V_{5}=V_{5\mathrm{L}}+\chi_{5} V_{2} V_{3}\;,
\end{align}
from which we define the nonlinear parts as $V_{4\mathrm{NL}} = \chi_{4}\left(V_{2}\right)^{2}$, $V_{5\mathrm{NL}} = \chi_{5} V_{2} V_{3}$. The coefficient $\chi_4$, for example, determines the coupling strength between the elliptic and quadrangular flow. Measurements of these couplings are of great interest as they are dynamically generated during the QGP expansion. They probe the transport and hadronization properties of the QGP, whose characterization is one of the main goals of high-energy heavy-ion collision experiments.

This paper establishes a new method to probe the nonlinear coupling of flow harmonics in the QGP expansion. We look at two collision systems at identical multiplicities but with small differences in their initial geometries. We study how the QGP responds to such differences in the dynamically-generated couplings between $V_2$, $V_3$, and higher-order harmonics $V_4$, $V_5$. We argue that the cleanest method to achieve this is to exploit isobars. These are nuclei with identical mass numbers but different deformations and radial profiles, providing the desired differences in collision geometries.  We focus on \ruru{} and \zrzr{} collisions, for which high-precision experimental data are available from the Relativistic Heavy Ion Collider (RHIC)~\cite{STAR:2021mii}. Using a transport model, we predict $\chi_{n}$ in these two systems, and systematically address the impact of nuclear structure on such observables. The $\chi_{n}$ show very small differences between \ruru{} and \zrzr{}, much smaller than the calculated differences for corresponding nonlinear terms $V_{n\mathrm{NL}}$. These small differences are ascribed to subleading nonlinear couplings, so far undetected in heavy-ion collisions, which are within the reach of the existing isobar data. We conclude that the different structures of two isobars can be exploited as a precision tool to access subtle nonlinear phenomena in the QGP expansion.

\paragraph*{Model, observables, goal.} 
We study \ruru{} and \zrzr{} collisions within the Glauber Monte Carlo model~\cite{Miller:2007ri}. The colliding ions are treated as collections of nucleons that are randomly distributed in each event according to a Woods-Saxon density,
\begin{align}\label{eq:2}
\rho(r,\theta,\phi)&\propto\frac{1}{1+e^{[r-R_0\left(1+\beta_2 Y_2^0(\theta,\phi) +\beta_3 Y_3^0(\theta,\phi)\right)]/a_0}},
\end{align}
with four nuclear structure parameters: nuclear diffusivity $a_0$, half-width radius $R_0$, quadrupole deformation $\beta_2$, and octupole deformation $\beta_3$. Each collision has a number of nucleons that participate in the interaction, which gives rise to the QGP. The evolution of QGP is modeled via the multi-phase transport code (AMPT) \cite{Lin:2004en}, version v2.26t5 in string-melting mode, which generates the hadrons in the final state. Observables are calculated using hadrons with $0.2<\pT<2$ GeV for events sorted in intervals of $\nch$: the charged hadron multiplicity for $\pT>0.1$ GeV and $|\eta|<0.5$, similar to the experiment~\cite{STAR:2021mii}. The observables of interest involve products of flow vectors $V_n=v_n e^{in\Psi_n}$ averaged over events with the same $\nch$, where $\Psi_n$ is the orientation of the harmonic and $v_n \equiv |V_n|$ is its amplitude. Such averages are computed within the framework of multi-particle correlations \cite{Bilandzic:2010jr,Bilandzic:2013kga}. In particular, we use a sub-event method \cite{Jia:2017hbm} by correlating particles in the pseudorapidity window $0<\eta<2$ with those having $-2<\eta<0$ to reduce the impact of non-collective (non-flow) correlations.

Our focus is on ratios of observables taken between \ruru{} and \zrzr{} collisions. We study the dynamical response of the QGP to small differences in the initial geometry induced by the nuclear structure. The $V_2$ and $V_3$ emerge as a response to the initial spatial deformations (or eccentricities) $\mathcal{E}_2$ and $\mathcal{E}_3$ \cite{Teaney:2010vd}, respectively. The response coefficient $K_n=V_n/\mathcal{E}_n$ is a probe of the medium properties, and has been investigated in model studies of isobar collisions \cite{Zhang:2021kxj,Jia:2021oyt,Nijs:2021kvn}. Here, we look instead at the coupling between harmonics represented by $\chi_n$. The advantage of such quantities is that they do not directly depend on eccentricities (Eq.~\eqref{eq:3}), and can be extracted directly from experimental $V_n$ data.

To this end, we continue from the decompositions in Eq.~\eqref{eq:1}. Defining $V_{n\mathrm{L}}$ as the vector that is uncorrelated with $V_{n\mathrm{NL}}$ in the considered event class~\cite{Yan:2015jma}, $\lr{V_{4\mathrm{L}}(V_2^{*})^{2}} = \lr{V_{5\mathrm{L}}V_2^{*}V_3^{*}}= 0$, one obtains an unique expression for the coupling coefficients,
\begin{align}
\label{eq:3}
\chi_4 = \frac{\langle V_4 (V_2^2)^*  \rangle}{\langle v_2^4 \rangle}, \hspace{30pt} \chi_5 = \frac{\langle V_5 (V_2 V_3)^*  \rangle}{\langle v_2^2 v_3^2 \rangle}.
\end{align}
It is clear both $\chi_4$ and $\chi_5$ are expressed in terms of quantities that are measurable. By construction, the $\chi_n$ encodes as well the effect of subleading couplings that go beyond the $V_2^2$ or $V_2 V_3$ terms. If the impact of such subleading couplings becomes more visible as we vary the initial geometry, then $\chi_n$ would change accordingly. The goal of this work is to expose and make a precision test of this feature using collisions of isobars, which permits us to claim evidence of subleading couplings in $V_4$ and $V_5$, driven by small differences in the initial geometry between \ruru{} and \zrzr{} collisions. 

The same could in principle be achieved by comparing other species at the same multiplicities, for instance, $^{238}$U+$^{238}$U versus $^{197}$Au+$^{197}$Au or $^{208}$Pb+$^{208}$Pb versus $^{129}$Xe+$^{129}$Xe, collisions that were already taken at RHIC and the Large Hadron Collider (LHC), respectively. However, as the expected contribution from subleading modes to $\chi_n$ is small, it might be beyond the genuine measurement systematics. In contrast, the isobar running mode guarantees that ratios of observables are devoid of measurement systematics~\cite{STAR:2019bjg}, as well as system dependence of final state effects~\cite{Zhang:2022fou}, which makes the extraction of novel nonlinear effects possible.
\begin{figure*}[t]
\includegraphics[width=0.89\linewidth]{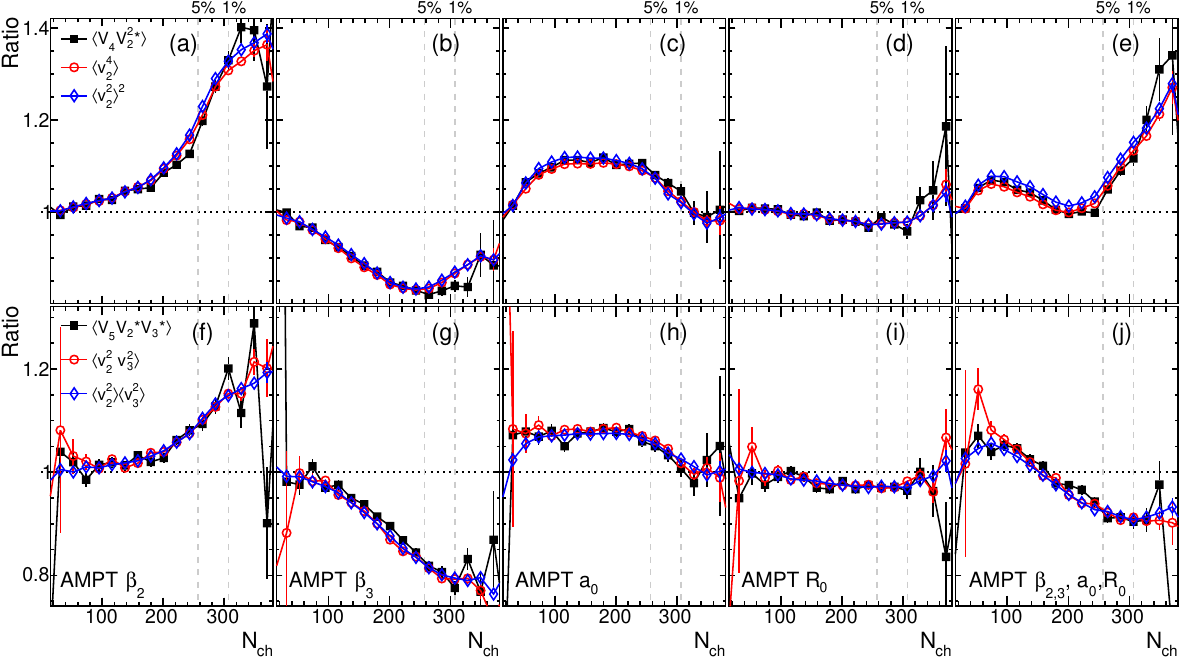}
\vspace*{-.4cm}
\caption{\label{fig:1} Isobar ratios of ${\lr{V_4\left(V_{2}^*\right)^{2}}}$, $\lr{v_2^4}$, $\lr{v_2^2}^2$ (top), and ${\lr{V_5 V_{2}^* V_3^*}}$, $\lr{v_2^2v_3^2}$, $\lr{v_2^2}\lr{v_3^2}$ (bottom). We consider the impact of $\beta_2$, $\beta_3$, $a_0$, $R_0$, and all of them are included, moving from left to right in the figure. The locations for 1\% and 5\% most central collisions are indicated by vertical dashed lines.}
\end{figure*}

\paragraph*{Nonlinear coupling coefficients.} 
To make progress, we perform the exercise recently proposed in Ref.~\cite{Zhao:2022uhl,Magdy:2022cvt}. We denote by $\mathcal{R}_\mathcal{O}$ the ratio of observable $\mathcal{O}$ taken between \ruru{} and \zrzr{} collisions at fixed $\nch$,
\begin{align}\label{eq:4}
\mathcal{R}_\mathcal{O} (\nch) = \frac{\mathcal{O}_{\mathrm{Ru+Ru}} (\nch)}{\mathcal{O}_{\mathrm{Zr+Zr}} (\nch)}.
\end{align}
We shall refer to such operation as the \textit{isobar ratio}. Taking the isobar ratio of the plane correlators $\langle V_4 (V_2^2)^* \rangle$ and $\langle V_5 V_2^*V_3^* \rangle$ yields,
\begin{align}\label{eq:5}
&\mathcal{R}_{\lr{V_4\left(V_{2}^*\right)^{2}}}=\mathcal{R}_{\chi_4} \mathcal{R}_{\lr{v_2^4}},\;\;\; \mathcal{R}_{\lr{V_5V_{2}^*V_{3}^*}}=\mathcal{R}_{\chi_5} \mathcal{R}_{\lr{v_2^2v_2^3}}.
\end{align}
The ratio at the same $\nch$ ensures a nearly perfect cancellation of final state effects for flow observables due to e.g. viscosity and hadronization~\cite{Zhang:2022fou}. One also in general expects that $\chi_n$ are nearly independent of initial-state features, i.e., $\mathcal{R}_{\chi_n}\approx 1$ \cite{Yan:2015jma}. Therefore, the ratio of the plane correlators should be given by the ratio of $\langle v_2^4 \rangle$ or $\langle v_2^2 v_3^2 \rangle$, which in turn are directly influenced by the nuclear structure parameters. To see this explicitly, we simulate generic high-statistics isobar collisions for several choices of Woods-Saxon densities listed in Tab.~\ref{tab:1}, which were found to describe ratios of several observables in isobar collisions~\cite{Jia:2021oyt}. These permit us to isolate the impact of individual features of the nuclear profiles. For instance, Case1/Case2 isolates the impact of $\beta_2$, while the combined effect of all four parameters is reflected by Case1/Case5.

Figure~\ref{fig:1} shows the $N_{\rm ch}$ dependence of the ratios in Eq.~\eqref{eq:5}, including the impact of the four Woods-Saxon parameters one at a time.  The top row shows results for $V_4$. Remarkably, we see that the ratio of plane correlators $\mathcal{R}_{\langle V_4 (V_2^2)^* \rangle}$ follows very precisely $\mathcal{R}_{\lr{v_2^4}}$, irrespective of the source of modifications in the nuclear structure induced to $v_2$. Within the precision of this plot, $\mathcal{R}_{\chi_n}=1$ is confirmed. This result reproduces nicely the preliminary measurement of this observable by the STAR collaboration~\cite{chunjian}. It is interesting to note that the $\mathcal{R}_{\langle v_2^2 \rangle^2}$ (open diamonds) also follows very closely the $\mathcal{R}_{\lr{v_2^4}}$, with a visible difference appearing only after a change in surface diffuseness, $a_0$. This behavior can be understood from the following identity,
\begin{align}\nonumber
&\mathcal{R}_{\lr{v_2^2}^2}-\mathcal{R}_{\lr{v_2^4}} =x(\mathcal{R}_{v_2\{4\}^4}-\mathcal{R}_{\lr{v_2^4}}),\hspace{5pt} x\equiv\! v_2\{4\}^4/\left(2\lr{v_{2}^2}^2\right),
\end{align}
where we have introduced the fourth-order cumulant of the fluctuations of $v_2$: $v_2\{4\}^4 = 2 \langle v_2^2 \rangle^2 - \langle v_2^4 \rangle$. In the top-middle panel of Fig.~\ref{fig:1}, a quick estimate gives $x\sim 0.1$ and $\mathcal{R}_{v_2\{4\}^4}\sim 1.35$ in mid-central collisions, which explains quantitatively why $\mathcal{R}_{\lr{v_2^2}^2}$ is about 2\% larger than $\mathcal{R}_{\lr{v_2^4}}$.

The lower panels of Fig.~\ref{fig:1} display similar comparisons for $V_5$. Our conclusions are unchanged. In this case, we predict in addition perfect agreement between the ratio of $\langle v_2^2 v_3^2 \rangle$ and that of $\langle v_2^2 \rangle \langle v_3^2 \rangle$ for all Woods-Saxon parameters. This is expected from the rather weak correlation between $v_2$ and $v_3$ in heavy ion collisions~\cite{Lacey:2013eia,Huo:2013qma,ATLAS:2015qwl,ALICE:2016kpq}.

\begin{table}[b]
\centering
\begin{tabular}{|l|cccc|}\hline 
   &\; $R_0$ (fm)\; & \;$a_{0}$ (fm)\;  & $\beta_{2}$ & $\beta_{3}$  \\\hline 
Case1 $^{96}$Ru & 5.09  & 0.46   & 0.162 & 0  \\
Case2          & 5.09  & 0.46   & 0.06  & 0  \\
Case3          & 5.09  & 0.46   & 0.06 & 0.20  \\
Case4          & 5.09  & 0.52   & 0.06 & 0.20  \\
Case5 $^{96}$Zr & 5.02  & 0.52   & 0.06 & 0.20  \\\hline\hline
$\vphantom{\displaystyle\frac{A_A}{A_A}}$ Ratios & \multicolumn{4}{|c|}{$\frac{\mathrm{\textstyle\small Case1}}{\mathrm{ \textstyle\small Case2}}$\;,\; $\frac{\mathrm{\textstyle\small Case2}}{\mathrm{ \textstyle\small Case3}}$\;,\;$\frac{\mathrm{\textstyle\small Case3}}{\mathrm{ \textstyle\small Case4}}$\;,\;$\frac{\mathrm{\textstyle\small Case4}}{\mathrm{ \textstyle\small Case5}}$\;,\;$\frac{\mathrm{\textstyle\small Case1}}{\mathrm{ \textstyle\small Case5}}$}\\\hline
\end{tabular}
\caption{\label{tab:1} Woods-Saxon parameters for $^{96}$Ru (Case1), $^{96}$Zr (Case5). The three other cases allow us to study the impact of nuclear structure on observables step by step. About 170 million AMPT events are generated for each case, about 10\% of the available real isobar data for one collision system.  The choice of parameters is motivated by previous comparisons between AMPT results and experimental data \cite{Jia:2021oyt}.} 
\end{table}

\begin{figure}[t]
\includegraphics[width=0.74\linewidth]{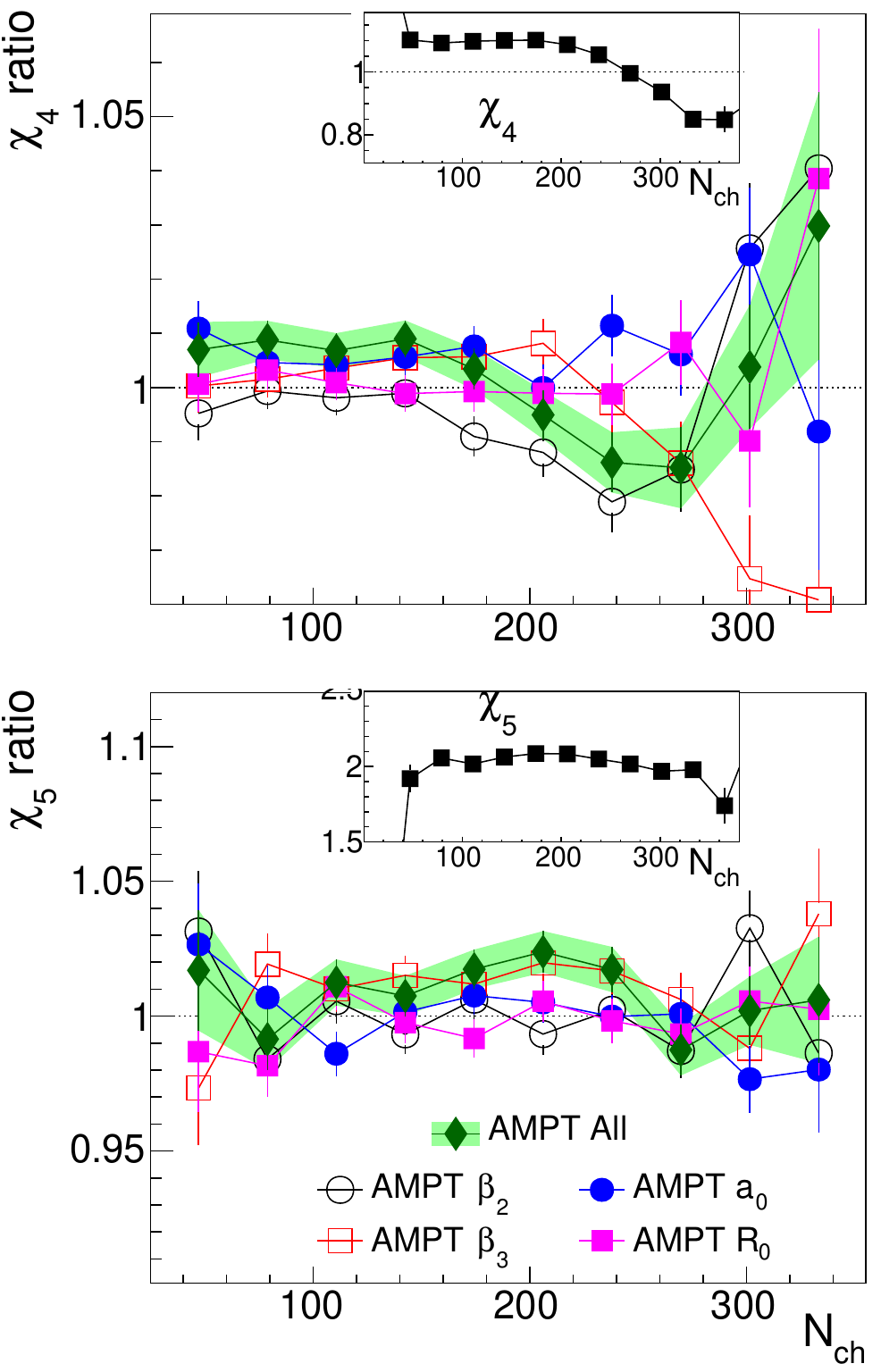}
\vspace*{-.4cm}
\caption{\label{fig:2} Isobar ratios of nonlinear response coefficients $\chi_4$ (top panel) and $\chi_5$ (bottom panel) as a function of $N_{\rm ch}$. The impact of various Woods-Saxon parameters is exhibited separately, as well as combined together. The insets show predictions for $\chi_4$ and $\chi_5$ averaged over all cases.}
\end{figure}

Having exposed the nonlinear couplings, we zoom in on these curves and examine the actual values of $\mathcal{R}_{\chi_n}$. Our predictions are given in Fig.~\ref{fig:2}. We do observe small systematic deviations of up to 1--2\% level induced by nuclear structure effects. Concerning $\chi_4$, the most significant deviations are associated with the nuclear deformation parameters $\beta_2$ and $\beta_3$. Larger $\beta_2$ in $^{96}$Ru leads to a slight decrease of $\chi_4$ from peripheral to central collisions, while the larger $\beta_3$ of $^{96}$Zr leads to the opposite effect. Similar effects are observed for $\chi_5$. The effects are largest in central collisions, where the impact of the $\beta_n$ are significant, and therefore any difference in the hydrodynamic response between isobars would be more visible.  Both ratios are, on the other hand, more weakly-dependent on the radial profile parameters, $a_0$ and $R_0$. 

The insets of Fig.~\ref{fig:2} show our predictions for $\chi_4$ and $\chi_5$ as a function of $\nch$. They exhibit a weak centrality dependence, with only a decrease in the 0--5\% centrality range, consistent with recent STAR measurement \cite{STAR:2020gcl} in \auau{} collisions. We stress that the precision reached in our analysis is in fact worse than what could be achieved based on the datasets available to the STAR collaboration. Therefore, percent level deviations could be easily detected with the isobar data.

This leads us to our main conclusion. The fact that $\mathcal{R}_{\chi_n} \neq 1$ implies the presence of additional nonlinear contributions to $V_4$ and $V_5$, driven by the nuclear structure effects. For instance, an important nonlinear contribution to $V_4$ beyond $V_2^2$ should be $V_3V_1$. It makes sense, then, that $\chi_4$ is affected by the large $\beta_3$ in $^{96}$Zr, as both $V_1$ and $V_3$ have a leading dependence on $\beta_3$ \cite{Jia:2021tzt}. However, since both $V_1$ and $V_3$ do not have a leading dependence on $\beta_2$,  the dependence of $\mathcal{R}_{\chi_4}$ on $\beta_2$ suggests another nonlinear contribution, potentially in the form of $V_3^2 V_2^*$~\footnote{In principle, the nonlinear mode $V_1^2V_2$ is also allowed. However, the $\pT$-integrated $V_1$ is known to be very small~\cite{ATLAS:2012at}. Thus, due to the presence of $V_1^2$, this mode should be strongly suppressed.}. A detailed study of such geometry-induced subleading couplings is beyond the scope of this paper. However, this analysis can be readily performed if our prediction is confirmed by the experimental measurements. We emphasize once more that, without resorting to isobar ratios, it would be impossible to achieve the precision required to detect such effects. This is a remarkable consequence of the isobar collision campaign, with profound implications for future precision QGP studies.

\paragraph*{Subleading nonlinear modes in $v_{4L}$.} 
In the framework of nonlinear coupling coefficients, the definition of the nonlinear term $V_{n\mathrm{NL}}$ (and therefore $\chi_n$) and the linear term $V_{n\mathrm{L}}$ are intertwined. This implies that, if $\chi_n$ coefficients are impacted by subleading nonlinear modes, such effect may also occur for $v_{n\mathrm{L}}$ and show up in the isobar ratio of $v_{n\mathrm{L}}$. We focus here on $n=4$.

\begin{figure*}[t]
\includegraphics[width=0.95\linewidth]{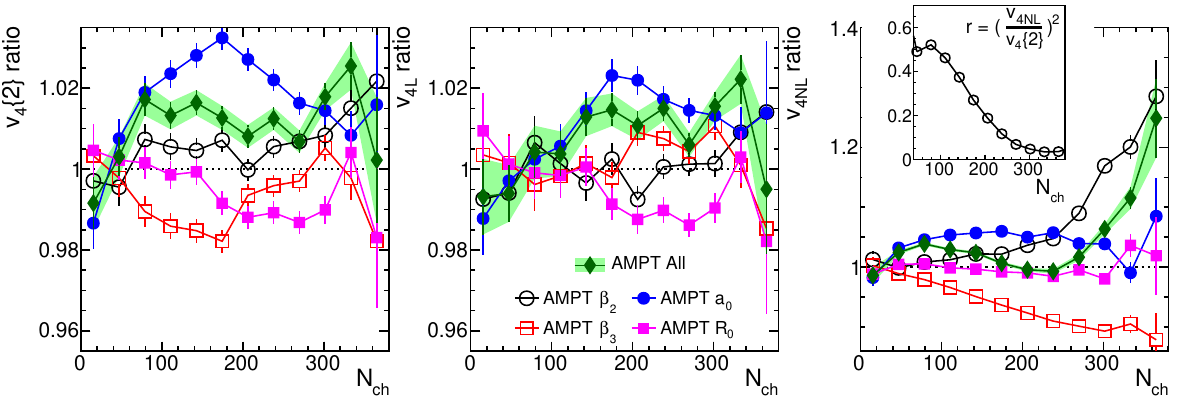}
\vspace*{-.4cm}
\caption{\label{fig:3} The ratios of quadrangular flow from a two-particle method, $v_4\{2\}$ (left), and the ratios of estimated linear component, $v_{\mathrm{4L}}$ (middle), and nonlinear component, $v_{\mathrm{4NL}} $ (right), as a function of $\nch$ in isobar collisions. Nuclear structure differences follow Table~\ref{tab:1}. The inset shows the fraction of $v_4^2$ carried by the nonlinear mode.}
\end{figure*}

The linear component can be isolated by combing Eq.~\eqref{eq:1} with a two-particle measurement of $V_4$, $v_4\{2\}^2\equiv \lr{v_4^2}$~\cite{ATLAS:2014ndd,ATLAS:2015qwl},
\begin{align}\label{eq:6}
v_{4\mathrm{L}}^2 = v_4\{2\}^2 -v_{4\mathrm{NL}}^2,\hspace{20pt} v_{\mathrm{4NL}}^2\equiv \chi_4^2\lr{v_2^4}.
\end{align}
The isobar ratios of the three quantities in the above equation are related via a simple identity,
\begin{align}\nonumber
 &\mathcal{R}_{v_4\{2\}^2} =\mathcal{R}_{v_{4\mathrm{L}}^2} + (\mathcal{R}_{v_{4\mathrm{NL}}^2}-\mathcal{R}_{v_{4\mathrm{L}}^2})r,\;\;\; r\equiv\!\frac{v_{4\mathrm{NL}}^2}{v_{4}\{2\}^2}\\\label{eq:7}
 &\mathcal{R}_{v_4\{2\}} \approx \mathcal{R}_{v_{4\mathrm{L}}} + (\mathcal{R}_{v_{4\mathrm{NL}}}-\mathcal{R}_{v_{4\mathrm{L}}})r
 \end{align}
The second line of the above equation is valid when all ratios are close to unity. Figure~\ref{fig:3} shows our predictions for the isobar ratio of $v_4$, as well as how the impact of nuclear structure manifests in the linear, $v_{\mathrm{4L}}$, and nonlinear mode, $v_{\mathrm{4NL}}$, extracted using Eq.~\eqref{eq:6}. The left panel displays the ratio $\mathcal{R}_{v_4\{2\}}$, which reveals small but rather complex dependencies on the nuclear structure parameters. The right panel shows the impact of such parameters on the nonlinear component, which we already observed from $\mathcal{R}_{\lr{v_2^4}}$ in the top panels of Fig.~\ref{fig:1} (however, note that $\mathcal{R}_{v_{4\mathrm{L}}}\approx \sqrt{\mathcal{R}_{\lr{v_2^4}}}$). The fraction of nonlinear contribution, shown in the inset, approaches zero in central collisions so that $v_{\mathrm{4NL}}$ impacts $\mathcal{R}_{v_4\{2\}}$ mostly in the non-central region. 

The middle panel of Fig.~\ref{fig:3} shows the ratio of linear modes, $\mathcal{R}_{v_{\mathrm{4L}}}$. We observe something important. The coefficient $\chi_4$ in Fig.~\ref{fig:2} shows a sensitivity to $\beta_2$ and $\beta_3$. The same does not occur for $v_{\mathrm{4L}}$, which can be understood from the fact that fourth-order eccentricity $\mathcal{E}_4$ does not depend on $\beta_2$ and $\beta_3$~\cite{Jia:2021tzt}. On the other hand, $v_{\mathrm{4L}}$ is more strongly impacted by the radial profile parameters, $a_0$ and $R_0$. These results imply that the effect of subleading modes to $v_{4\mathrm{L}}$ is complementary to that observed for $v_{4\mathrm{NL}}$. Both observables probe such phenomena but respond to initial-state deformations in different ways. We stress, once more, that all these features can be accessed experimentally via high statistics isobar data.

\paragraph*{Conclusion \& Prospects.} Detailed transport simulations predict that the precision reached in the isobar ratios of higher-order flow harmonic coefficients permits one to scrutinize the effect of subleading couplings to $V_4$ and potentially $V_5$. If this prediction is confirmed in experiments, it will establish isobar collisions as a unique tool to study subleading nonlinear effects in the QGP expansion, exposed by the different structures of the two isobars. The largest subleading modes to $V_4$ should come from $V_1V_3$ and $V_3^2 V_2^*$, namely,
\begin{equation}
V_4 = V_{4\mathrm{L}} + \chi_{4,22} V_2^2 + \chi_{4,31} V_1 V_3 + \chi_{4,332} V_3^2 V_2^*.
\end{equation}
A self-consistent framework to define both $V_{4\mathrm{L}}$ and the coupling coefficients in the presence of multiple nonlinear components has been derived in Ref.~\cite{Giacalone:2018wpp}, and could be readily applied to isobar data analysis. So far, subleading couplings have only been discussed in the analysis of $V_6$, $V_7$, $V_8$ in \pbpb{} collisions at LHC energies. The rich nonlinear structure of, e.g., $V_6$ is, however, difficult to fully clarify via isobar ratios in \ruru\ and \zrzr\ collisions at RHIC, as multiplicities are not high enough. If larger isobar pairs are collided in the future, potentially at higher collision energies at the LHC, such a possibility will be realized. Note that identification of nonlinear effects via isobar ratios allows cancellation of model uncertainties, which is not possible by looking at one collision system such as \pbpb{} at the LHC. Therefore, isobar collisions will permit us to access experimentally the full intricacies of the nonlinear hydrodynamic response, thus opening a new promising opportunity for data-driven determinations of the QGP transport properties.

\paragraph*{Acknowledgments.} This research of J.J and C.Z is supported by DOE DE-FG02-87ER40331. The research of G.G. is funded by the Deutsche Forschungsgemeinschaft (DFG, German Research Foundation) under Germanys Excellence Strategy EXC2181/1-390900948 (the Heidelberg STRUCTURES Excellence Cluster), within the Collaborative Research Center SFB1225 (ISOQUANT, Project-ID 273811115). We acknowledge Somadutta Bhatta for their useful comments.
\bibliography{deform}{}
\end{document}